\documentclass{elsart5p}

\usepackage{graphics}
\usepackage{epsfig}
\usepackage{xcolor}
\usepackage{subfigure}
\usepackage{subcaption}
\usepackage{color}

\journal{Physics Letters B}

\begin{document}

\renewcommand{\textfraction}{0.00000000001}
\renewcommand{\floatpagefraction}{1.0}
\begin{frontmatter}
\title{Helicity dependent cross sections for the photoproduction of $\pi^0\pi^{\pm}$ pairs from quasi-free nucleons}

\author[Basel]{D.~Ghosal \corauthref{*}}\corauth[*] {Corresponding author, email address: dghosal@liverpool.ac.uk, Present affiliation: University of Liverpool, UK},
\author[Mainz]{V.~Sokhoyan},
\author[Tomsk]{A.~Fix},
\author[Basel]{S.~Lutterer},
\author[Basel]{S.~Abt}, 
\author[Mainz]{P.~Achenbach},
\author[Bonn]{F.~Afzal},
\author[Regina]{Z.~Ahmed},
\author[Glasgow]{J.R.M.~Annand},          
\author[York]{M.~Bashkanov},
\author[Bonn]{R.~Beck},         
\author[Mainz]{M.~Biroth},
\author[Dubna]{N.~Borisov},
\author[Pavia]{A.~Braghieri},           
\author[Washington]{W.J.~Briscoe},           
\author[Mainz]{F.~Cividini},
\author[Halifax]{C.~Collicot},
\author[Pavia,PaviaUni]{S.~Costanza},         
\author[Mainz]{A. Denig},
\author[Basel]{M.~Dieterle},             
\author[Dubna]{A.S.~Dolzhikov},
\author[Washington]{E.J.~Downie},            
\author[Mainz]{P.~Drexler}, 
\author[York]{S.~Fegan},
\author[Glasgow]{S.~Gardner}, 
\author[Glasgow]{D.I.~Glazier},
\author[Dubna]{I.~Gorodnov},
\author[Mainz]{W.~Gradl},
\author[Basel]{M.~G{\"u}nther},
\author[Moscow]{D.~Gurevich},
\author[Mainz]{L.~Heijkenskj{\"o}ld},
\author[Sackville]{D.~Hornidge}, 
\author[Regina]{G.M.~Huber},
\author[Basel]{N.~Jermann},  
\author[Basel]{A.~K\"aser},         
\author[Lebedev,Mainz]{V.L.~Kashevarov},         
\author[Zagreb]{M.~Korolija},  
\author[Basel]{B.~Krusche},
\author[Dubna]{A.~Lazarev},
\author[Glasgow]{K.~Livingston}, 
\author[Glasgow]{I.J.D.~MacGregor},       
\author[Kent]{D.M.~Manley},
\author[Mainz]{P.P.~Martel},
\author[Basel]{Ch.~Meier},
\author[Amherst]{R.~Miskimen},
\author[York]{M.~Mocanu},
\author[Mainz]{E.~Mornacchi},
\author[Glasgow]{C.~Mullen},
\author[Dubna]{A.~Neganov},
\author[Mainz]{A.~Neiser},
\author[Mainz]{M.~Ostrick},
\author[Mainz]{P.~Otte},
\author[Regina]{D.~Paudyal},
\author[Pavia]{P.~Pedroni}, 
\author[Glasgow]{A.~Powell},           
\author[Bochum]{G.~Reicherz},
\author[Basel]{T.~Rostomyan}, 
\author[Bonn]{K.~Spieker},
\author[Mainz]{O.~Steffen},
\author[Washington]{I.I.~Strakovsky},        
\author[Bonn]{A.~Thiel},      
\author[Mainz]{M.~Thiel},        
\author[Mainz]{A.~Thomas},              
\author[Mainz]{M.~Unverzagt}, 
\author[Dubna]{Yu.A.~Usov},
\author[Mainz]{S.~Wagner},
\author[Basel]{N.K.~Walford},                      
\author[York]{D.P.~Watts},
\author[Basel]{D.~Werthm\"uller},
\author[Mainz]{J.~Wettig},
\author[Basel]{L.~Witthauer},
\author[Mainz]{M.~Wolfes},
\author[York]{N.~Zachariou}

\address[Basel] {Department of Physics, University of Basel, Ch-4056 Basel, Switzerland}
\address[Mainz] {Institut f\"ur Kernphysik, University of Mainz, D-55099 Mainz, Germany}
\address[Tomsk] {Tomsk Polytechnic University, Tomsk, Russia}
\address[Bonn] {Helmholtz-Institut f\"ur Strahlen- und Kernphysik, University Bonn, D-53115 Bonn, Germany}  
\address[Regina] {University of Regina, Regina, SK S4S-0A2 Canada}
\address[Glasgow] {SUPA School of Physics and Astronomy, University of Glasgow, G12 8QQ, United Kingdom}
\address[York] {Department of Physics, University of
York, Heslington, York, Y010 5DD, UK}
\address[Dubna] {Joint Institute for Nuclear Research, 141980 Dubna, Russia}
\address[Pavia] {INFN Sezione di Pavia, I-27100 Pavia, Pavia, Italy}
\address[PaviaUni] {Dipartimento di Fisica, Universit\`a di Pavia, Pavia, Italy}
\address[Washington] {Center for Nuclear Studies, The George Washington University, Washington, DC 20052, USA}
\address[Halifax] {Department of Astronomy and Physics, Saint Mary's University, E4L1E6 Halifax, Canada}
\address[Moscow] {Institute for Nuclear Research, RU-125047 Moscow, Russia}
\address[Sackville] {Mount Allison University, Sackville, New Brunswick E4L1E6, Canada}
\address[Lebedev] {Lebedev Physical Institute, RU-119991 Moscow, Russia}
\address[Zagreb] {Rudjer Boskovic Institute, HR-10000 Zagreb, Croatia}
\address[Kent] {Kent State University, Kent, Ohio 44242, USA}
\address[Bochum] {Institut f\"ur Experimentalphysik, Ruhr Universit\"at, 44780 Bochum, Germany}
\address[Amherst] {University of Massachusetts, Amherst, Massachusetts 01003, USA}

\begin{abstract}
Photoproduction of $\pi^0\pi^{\pm}$-pairs from quasifree nucleons bound in the deuteron has been
investigated to study the helicity dependence of this reaction. Measurements with a liquid
deuterium target were used to extract the unpolarized cross sections for reactions on protons and neutrons. A deuterated, longitudinally polarized solid-butanol target, together with a circularly polarized photon beam, determined the double polarization observable $E$. From these results the spin-dependent cross sections $\sigma_{1/2}$ and $\sigma_{3/2}$, corresponding to the anti-parallel and parallel spin configurations of the beam photon and target nucleon, have been derived. The measurements were performed at the Mainz MAMI accelerator with tagged, circularly-polarized photon beams produced via bremsstrahlung from longitudinally polarized electron beams. The reaction products were detected with an almost $4\pi$ solid-angle covering calorimeter composed of the Crystal Ball and TAPS detectors, supplemented by plastic scintillation detectors for charged particle identification. The results are sensitive to sequential decays of nucleon resonances via intermediate states and also to the decay of nucleon resonances by emission of charged $\rho$ mesons, and are compared to recent model results.
\end{abstract}
\end{frontmatter}

\section{Introduction}
The basic properties of the fundamental forces of nature are reflected
in the excitation spectrum of composite objects formed by them. The excitation spectrum of 
hadrons, such as protons and neutrons, is closely related to the properties of the strong 
interaction in the energy regime where it cannot be described by perturbative methods. 
The experimental approach to study the excitation spectrum of such systems is in principle
similar to the detailed study of the electromagnetic interaction with the spectroscopy of 
atomic levels by optical methods or the spectroscopy of nuclear levels by the detection of 
$\gamma$-radiation emitted by their decays. However, in practice the study of excited nucleon 
states is complicated by the fact that they decay by the emission of mesons via 
the strong interaction, resulting in very short life times and hence the states have large widths.
The states are therefore not well separated in energy like atomic or nuclear levels, but generally overlap. The original database for nucleon resonances came from elastic pion scattering. However, during the last two decades photon induced reactions have made important contributions~\cite{Thiel_22}. Photoproduction of mesons is a very versatile tool for the study of the strong interaction in different contexts. The impact of such reactions on the excitation spectrum of
the nucleon is visible in the Review of Particle Physics~\cite{PDG_20}, where for many excited nucleon states results from photoproduction reactions have become decisive in determining their properties.

The present experiment addresses three important topics in this field. Without the measurement 
of polarization observables, it is impossible to disentangle the nucleon excitation spectrum. 
Partial wave analysis methods that can identify states of different quantum numbers need as 
input not only differential cross sections, but also polarization observables, which are 
particularly sensitive to interference terms in the amplitudes. For the production of single 
pseudoscalar (PS) mesons~\cite{Chiang_97}, at least beam, target, and recoil single-polarization and 
four carefully chosen double polarization observables must be measured for a `complete' 
experiment. The double polarization observable $E$ discussed in this work, which quantifies 
the ratio of contributions with photon and nucleon spin parallel or antiparallel, is an important 
one~\cite{Barker_75}. It allows contributions from nucleon resonances with spin $J=1/2$ and $J\geq 3/2$ to be separated.
This is not only true for single meson production, but also for the production of pion pairs.

The production of pairs of PS mesons is important for several reasons. 
The most obvious one is that one expects that nucleon states with more complicated excitations,
in particular those which have both elementary oscillators excited \cite{Thiel_15} in the picture 
of a constituent quark model, decouple from direct decays to the nucleon ground state and tend to
decay via cascades involving an intermediate state in which only one oscillator de-excites.
Such decays result in the emission of meson pairs. The restriction to single meson production 
reactions is therefore strongly biased against contributions from such states. Furthermore,
the study of pion pairs gives also access to resonance decays involving heavier mesons that
decay to them. These are in particular the $\sigma$ and $\rho$ mesons ($f_0(500)$ and $\rho(770)$
\cite{PDG_20}). The first is best studied in the $\pi^0\pi^0$ final state, but the latter, 
due to isospin conservation, contributes only to $\pi^+\pi^-$ and to $\pi^0\pi^{\pm}$.

So far, mostly photoproduction off protons has been studied. However, since the
electromagnetic interaction does not conserve isospin, reactions with neutron targets must also
be investigated to determine the isospin dependence of the reaction mechanisms \cite{Krusche_11}.
This can only be done with the use of light nuclear targets, such as deuterium.
However, using nuclear targets introduces complications as not only the produced mesons, but also the recoil nucleons, must be identified and effects of Fermi motion and 
Final State Interaction (FSI) processes must be considered in the interpretation of the results.

Photoproduction of mixed charged (and doubly charged) pion pairs is also important for a different topic, namely the study of in-medium properties of hadrons. It has been much discussed in the 
literature, in particular in the context of heavy ion reactions, that properties of hadrons may 
change in dense and/or hot nuclear matter (see e.g. \cite{Leupold_10}), in the extreme leading to 
partial chiral symmetry restoration. However, even for nuclear matter at normal density, effects on 
the mass and/or width of hadrons have been suggested. An experimentally observed (see e.g. 
\cite{Bianchi_93}), but never understood effect is the almost complete suppression of the 
second and third resonance bumps in total photoabsorption for nuclear targets. These structures,
which are prominent for the free nucleon in the $\gamma N\rightarrow X$ reaction, are not observed 
for nuclei with mass numbers larger than $A=4$. Total photoabsorption is an inclusive reaction 
that does not relate this effect to any specific reactions, but by using exclusive 
production processes the bump-like structure of the second resonance peak can be traced to the 
production of pion pairs \cite{Krusche_99,Krusche_03,Krusche_04}. Pairs with at least one charged 
pion are of interest, not only because the production cross section is large, but also because they 
can result from the decay of the $\rho(770)$ meson. The latter is one of the prime candidates for 
strong in-medium modifications of hadrons \cite{Leupold_10}. Such effects (broadening, mass shifts) 
would also influence nucleon resonances that decay by its emission \cite{Post_04} and could 
contribute to the suppression of the higher resonance regions in total photoabsorption.
  
\section{Photoproduction of mixed-charge pion pairs}
The reaction formalism and the sets of observables for the photoproduction of PS
meson pairs are discussed in \cite{Roberts_05,Fix_11} and a field theoretic description of the 
two-pion production process is given in \cite{Haberzettl_19}. Photoproduction of PS meson pairs 
requires the measurement of eight observables as a function of five kinematic parameters to 
determine just the magnitudes of the amplitudes, and 15 observables have to be measured to also fix 
their phases \cite{Roberts_05}. Such an experiment is impractical, but even the 
measurement of a few polarization observables in addition to the unpolarized cross sections can
give valuable input for reaction models.    
 
The first precise experimental results for total cross sections and pion-pion, pion-nucleon 
invariant mass distributions for the mixed-charge state of pion pairs revealed strong 
discrepancies compared to model predictions. The DAPHNE experiment measured the quasi-free $\gamma n\rightarrow p\pi^0\pi^-$ reaction \cite{Braghieri_95,Zabrodin_99} and TAPS first 
investigated the $\gamma p\rightarrow n\pi^0\pi^+$ final state \cite{Langgaertner_01}. 
Both experiments were carried out at the MAMI accelerator and covered energies up to the maximum of 
the second resonance region. The total cross sections in the second resonance region were 
significantly larger and the invariant-mass distributions had different shapes compared to the available models~\cite{Gomez_96}. The invariant mass distributions of the pion-pion pairs
suggested a significant contribution from decays of the $\rho$-meson, which was previously ignored in the models for this energy range because of its large nominal mass (770~MeV). However, possible contributions from the low-energy tail of its mass distribution could be due to decays of the $N(1520)3/2^-$ resonance into $N\rho$, as discussed in \cite{Ochi_97}, or due to off-shell $\rho$-mesons in Kroll-Rudermann like diagrams (see Ref.~\cite{Krusche_03} for a summary). The latter mechanism has almost no effects on the other isospin channels, because the production of uncharged $\rho^0$ mesons in Kroll-Rudermann diagrams is suppressed. 

Apart from total cross sections and invariant mass distributions, the helicity split
of the $\gamma n\rightarrow p\pi^0\pi^-$ reaction \cite{Ahrens_11} was measured with the DAPHNE experiment up to the maximum of the second resonance region. The result was a dominant contribution of the $\sigma_{3/2}$ component (photon and nucleon spin aligned, see below), which would fit to the excitation of the $N(1520)3/2^-$ resonance. The helicity split was also measured with the DAPHNE setup for the reaction $\gamma p\rightarrow n\pi^0\pi^+$ with a free proton target~\cite{Ahrens_03}.

The beam-helicity asymmetry $I^{\odot}$, measured with a circularly polarized photon beam,
was reported for the $\gamma p\rightarrow n\pi^0\pi^+$ reaction off free protons up
to the second resonance region \cite{Krambrich_09,Zehr_12} and for quasifree nucleons 
bound in the deuteron up to photon energies of 1.4~GeV \cite{Oberle_14}. The comparison
to model predictions revealed large, non-understood discrepancies, particularly at low
incident photon energies.
  
\section{Experiment}
\label{sec:setup}
In the present experiment, the helicity-split of the cross section for the reactions
$\gamma p\rightarrow n\pi^0\pi^+$ and $\gamma n\rightarrow p\pi^0\pi^-$ off quasifree
nucleons bound in the deuteron was measured for photon energies up to 1.4~GeV. 

The primarily measured polarization observable in the notation of Ref.~\cite{Roberts_05}
is the asymmetry $P_z^{\odot}$, which is identically defined as the better known observable 
$E$ for single meson photoproduction by
\begin{equation}
E=\frac{\sigma_{1/2}-\sigma_{3/2}}{\sigma_{1/2}+\sigma_{3/2}}=\frac{\sigma_{1/2}-\sigma_{3/2}}{2\sigma_{0}}\;,
\label{eq:e}
\end{equation} 
For a circularly polarized photon beam and a longitudinally polarized target, two different 
relative spin orientations, parallel or antiparallel, corresponding to the cross sections 
$\sigma_{3/2}$ $(\uparrow\uparrow)$ and $\sigma_{1/2}$ $(\uparrow\downarrow)$ are possible, 
which are termed helicity-3/2 and helicity-1/2. These two configurations correspond to the 
excitation of nucleon resonances to the electromagnetic couplings $A_{3/2}$ and $A_{1/2}$ 
\cite{PDG_20}. Nucleon resonances with spin $J$=1/2 contribute only to $\sigma_{1/2}$ while 
those with $J\geq 3/2$ (or non-resonant backgrounds) can contribute to both. 

The experiment required longitudinally polarized deuterium nuclei as a target. These were 
provided by a solid deuterated-butanol target (C$_{4}$D$_{9}$OD) of the frozen-spin type 
\cite{Rohlof_04}. Average polarization of the deuterons was between 55\% and 62\%. 
The polarization was 
corrected for the quasi-free nucleons bound in the deuteron for the d-wave component
of the deuteron wave function ($\approx$6\% correction), resulting in the effective polarization
degree $P_{T}$. 
 
The background arising from reactions with the unpolarized heavy nuclei (mainly carbon, 
small amounts of oxygen) in the butanol molecules was determined with additional measurements 
using a carbon-foam target with the same geometry (target length and radius 2.0~cm) and the 
same surface density for heavy nuclei as the butanol target and a liquid deuterium target 
(length 4.72~cm, radius 4~cm). 

\subsection{Experimental setup} 
The data used for the present analysis have already been analyzed for the reactions 
$\gamma N\rightarrow N\pi^0$ \cite{Dieterle_17}, $\gamma N\rightarrow N\eta$ 
\cite{Witthauer_16,Witthauer_17}, $\gamma N\rightarrow N\pi\eta$ \cite{Kaeser_18},
and $\gamma N\rightarrow N\pi^0\pi^0$ \cite{Dieterle_20}. This data set is very well
understood and all relevant experimental details are given in the respective publications, therefore only a short summary is given here.

The experiment was performed at the MAMI accelerator in Mainz \cite{Kaiser_08} with the
Glasgow tagged photon facility \cite{McGeorge_08} using a circularly polarized photon beam. 
The photon beam was prepared with bremsstrahlung from a longitudinally polarized electron beam 
penetrating a copper radiator of 10 $\mu$m thickness. Typical values of the electron polarization $P_{e^-}$ were $\approx$\,80\%. The energy-dependent circular polarization 
$P_{\odot}$ of the photon beam followed from the polarization transfer formula given in 
Ref.~\cite{Olsen_59}
\begin{equation}
P_{\odot} = P_{e^{-}}\cdot(4x-x^{2})/(4-4x+3x^{2})~,\label{eq:olsen}
\end{equation} 
with $x=E_{\gamma}/E_{e^-}$, where $E_{e^-}$, $E_{\gamma}$ are electron and gamma beam energies,
respectively.

The targets were placed in the center of the main detector, the Crystal Ball (CB) 
\cite{Starostin_01} composed of 672 NaI(Tl) crystals, covering 20$^\circ$ to 160$^\circ$ of 
the polar angle in a spherical geometry supplemented at forward angles from 5$^\circ$ to 
20$^\circ$ by a hexagonal wall of 366 BaF$_2$ crystals from the TAPS detector \cite{Gabler_94}. 
The target was surrounded by a cylindrical particle identification detector (PID) comprising 
24 plastic scintillator strips, covering 15$^\circ$ in the azimuthal angle each \cite{Watts_05}.
Identification of charged particles hitting the TAPS detector was done with individual,
hexagonal plastic scintillators in front of each crystal. In total, the setup covered about
97\% of the total solid angle and could detect photons (from decays of neutral mesons), charged
pions, and recoil protons and neutrons.
 
\section{Data Analysis}
\subsection{Particle and Reaction Identification}
The reaction identification was almost identical to that in Ref.~\cite{Oberle_14}. 
In the first step, hits in the main calorimeters CB and TAPS were
assigned as `charged' or `neutral' depending on the response of the charged particle
detectors in front of them. For the analysis of the quasifree 
$\gamma d\rightarrow p\pi^0\pi^-(p)$ reaction, events with two charged (hypothesis $p$, $\pi^-$)
and two neutral (photons from $\pi^0$ decay) hits were accepted. 
For $\gamma d\rightarrow n\pi^0\pi^+(n)$, events with one charged ($\pi^+$) and three neutral 
(hypothesis $n$, $2\gamma$) were analyzed. For both reactions, the nucleon in brackets was the
undetected spectator nucleon. In rare cases, also the spectator nucleon can be detected due to
extreme Fermi momenta. This was not explicitly analyzed but included in the simulation of
detection efficiency with realistic Fermi momentum distributions.  

\begin{figure}[h!]
  \centerline{\resizebox{0.5\textwidth}{!}{
    \includegraphics[width=0.45\textwidth]{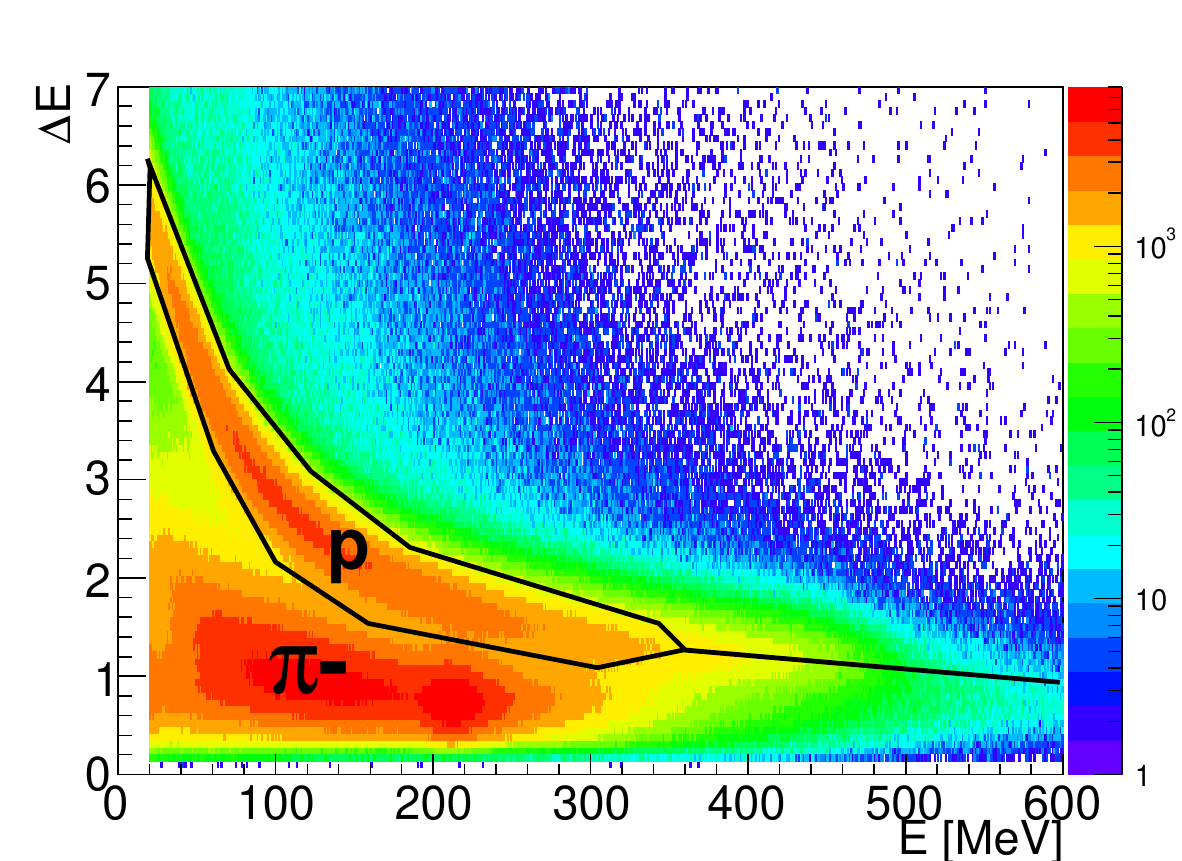}
    \includegraphics[width=0.45\textwidth]{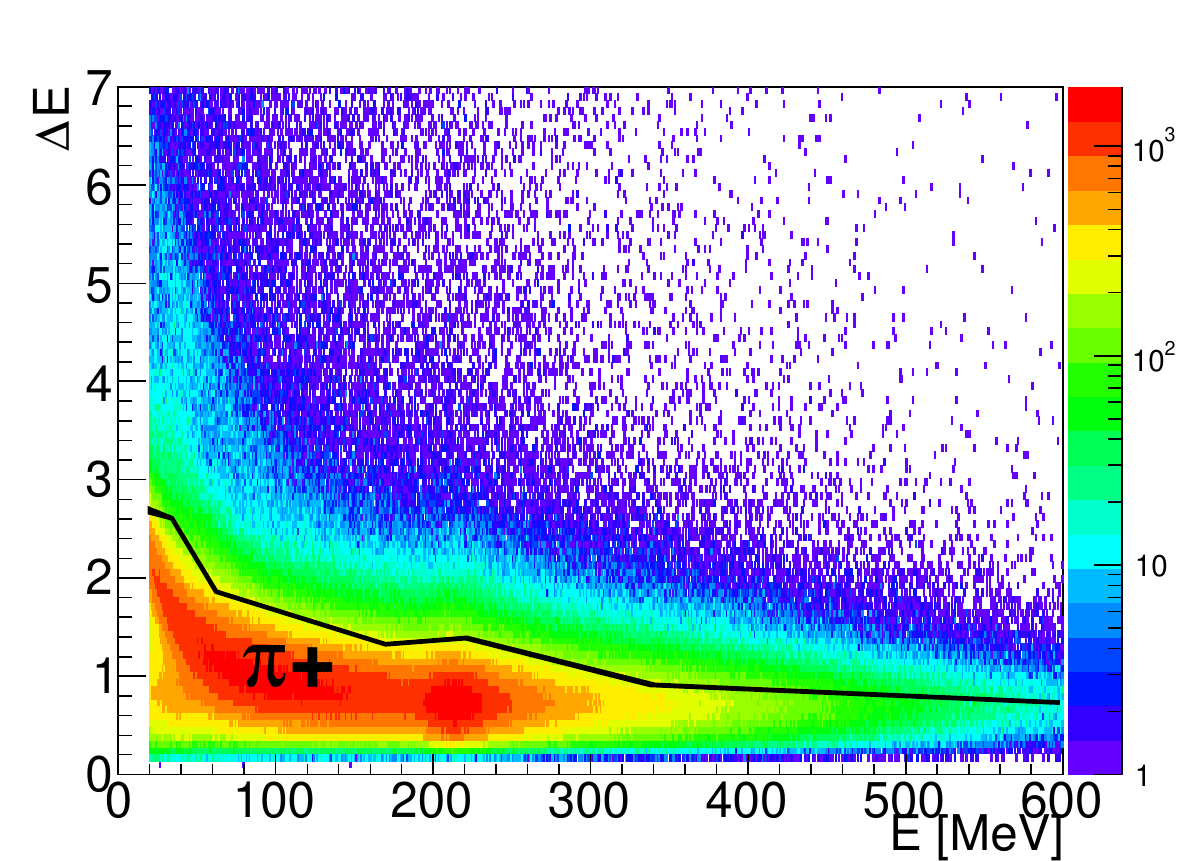}
}}
\caption{Identification of protons and charged pions in CB with a $\Delta E-E$ analysis.
Left hand side: events with two charged particles and two neutrals ($p\pi^-\pi^0$), 
Right hand side: events with one charged particle and three neutrals ($n\pi^+\pi^0$).} 
\label{fig:discri} 
\end{figure}

In the next step, calorimeter hits were assigned to photons, charged pions, protons, and neutrons. Charged pions
and protons hitting the CB were distinguished by an $E-\Delta E$ analysis comparing the energy
deposition in the CB and the energy loss in the PID. This is shown in Fig.~\ref{fig:discri}.   

For hits in the TAPS detector `charged' and `neutral' can be assigned due to the
response of the scintillation detector for charged particles. In addition, the pulse-shape
analysis of the response of the BaF$_2$ crystals separates photons and neutrons. Time-of-flight
versus energy analysis provides an additional separation of protons, photons, and neutrons. 

\begin{figure}[thb]
\centerline{\resizebox{0.5\textwidth}{!}{
  \includegraphics[width=0.45\textwidth]{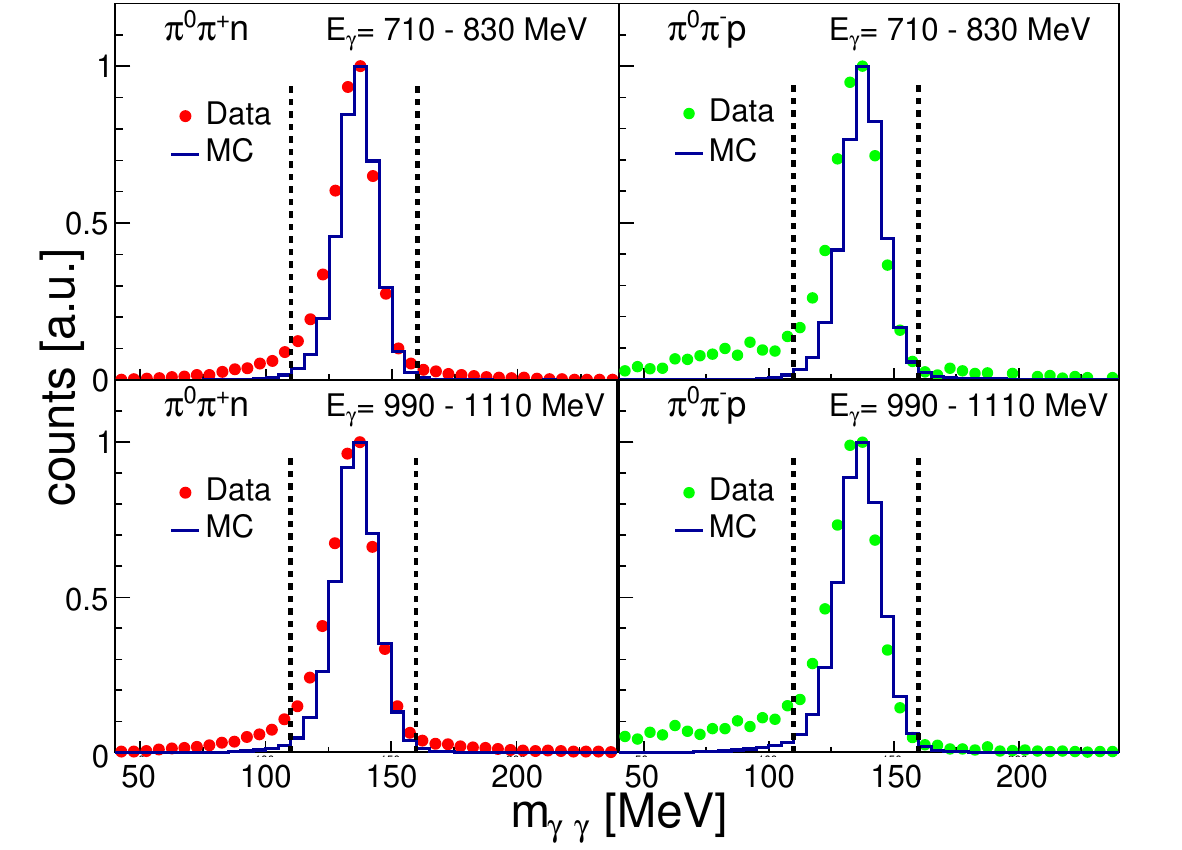}
}}
\caption{Invariant mass spectra for two typical energy ranges indicated in the figure. 
Left hand side:  $\pi^{0}\pi^{+}n$ final state, 
Right hand side: $\pi^{0}\pi^{-}p$ compared to MC simulations of the same reactions. 
The dashed vertical lines indicate cuts applied for further analysis.}
\label{fig:invmass} 
\end{figure}

The only missing information at this stage is the separation of neutrons and photons hitting the CB due to too short flight pass for time-of-flight analysis and the 
identification of charged pions in TAPS, since all usable methods like $E-\Delta E$ have too low resolution to exclude protons from background reactions. Therefore, events with charged pions in TAPS (polar angles $\leq 20^{\circ}$) were discarded and approximately corrected in the detection efficiency simulations. 

For events with three neutral hits in the CB, neutrons were distinguished from photons using a $\chi^{2}$ analysis of all possible combinations to identify a pair of photons arising from a $\pi^{0}$ decay. The $\chi^2$ of all hypothetical photon pairs was calculated from   
\begin{equation}
\chi^2 = \frac{(m_{\gamma\gamma}(k)-m_{\pi^0})^2}{\Delta m_{\gamma\gamma}(k)},\;\; 
\end{equation}
where k~=~1,2,3, $m_{\gamma\gamma}(k)$ are the invariant masses of the three possible combinations of neutral 
hits to pion-decay photons, $\Delta m_{\gamma\gamma}(k)$ are their uncertainties, and $m_{\pi^0}$ 
is the nominal pion mass. The pair with the best $\chi^2$ was chosen for the $\pi^0$ photons and 
the third neutral hit was assigned to the neutron. 

Invariant mass spectra for the two reactions and two typical incident photon energies are presented in Fig.~\ref{fig:invmass}. The experimental results for both reactions are compared to the line shapes generated by a Monte Carlo (MC) simulation using the Geant4 code \cite{Geant4}. They agree reasonably well with the experimental data and the level of background is low. The lower level of 
background at small invariant masses for the $\pi^0\pi^{+}n$ final state is due to the preceding $\chi^2$ analysis. The nominal invariant mass $m_{\pi^0}$ of the $\pi^0$ meson was used, as in Ref.~\cite{Oberle_14}, to improve the experimental resolution in further data analysis steps.

\begin{figure}[thb]
\centerline{\resizebox{0.5\textwidth}{!}{
  \includegraphics[width=0.45\textwidth]{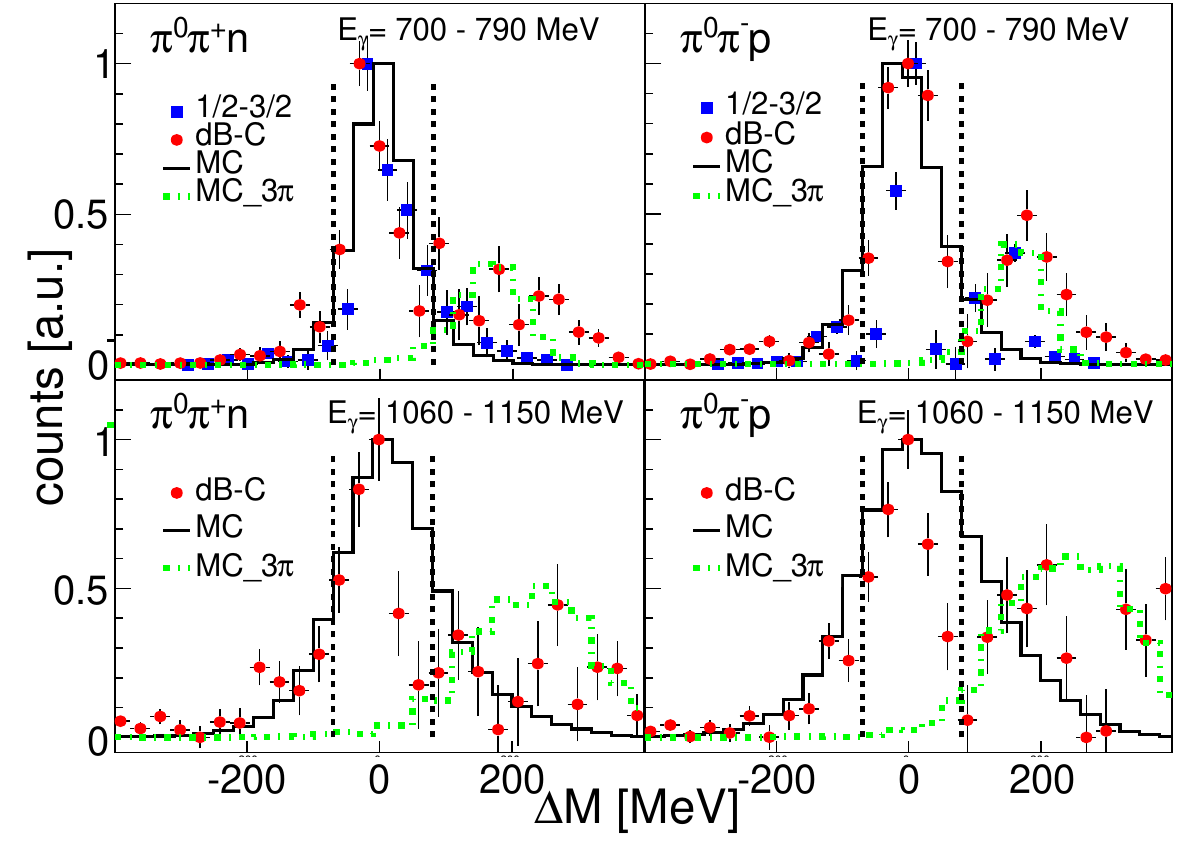} 
}}
\caption{Missing mass spectra for two typical energy ranges indicated in the figure. 
Plotted are the butanol data after subtraction of the carbon background (red circles).
For the lower energy range also the difference of the count rates for the two helicity states 
(blue squares). 
Left hand side:  $\pi^{0}\pi^{+}n$ final state, 
Right hand side: $\pi^{0}\pi^{-}p$.
Both reactions are compared to line shapes from MC simulations (black lines) which include background from three-pion production processes (green dash-dotted lines). Lastly, the vertical dotted lines indicate cuts
applied for further analysis.}
\label{fig:mismas} 
\end{figure}

The final step of event identification required the suppression of background from other reactions 
that have a neutral, a charged pion, and a recoil nucleon in the final state but also have additional undetected particles. Of particular concern are triple-pion production processes (including the 
$\gamma N\rightarrow N\eta\rightarrow N\pi^0\pi^+\pi^-$ reaction) where one charged pion can be lost
or misidentified as a recoil proton (in events where the recoil neutron was missed). 
   
The simplest condition to be fulfilled is `coplanarity', that is the azimuthal angles of the momenta
of the recoil nucleon and the two-pion system must be back-to-back in the laboratory frame. This
condition was tested as in \cite{Oberle_14}, the observed background level was small and some 
background was removed with a $2\sigma$ cut around the 180$^{\circ}$ nominal peak. However, this condition is 
not very stringent because undetected charged pions with low momenta do not significantly influence the angular balance. 

More effective is a missing mass analysis for which the recoil nucleon, although detected, is treated
as a missing particle and its mass is calculated from the kinematic parameters of the meson pair by
\begin{equation}
 \Delta M =\ \left | P_{\gamma}\ +\ P_N\ -\ P_{\pi^0}\ -\ P_{\pi^{\pm}}\right | - m_N,
\end{equation}
where $P_\gamma$, $P_{N}$ are the four-vectors of incident photon and incident nucleon (assumed to be
at rest, with the distribution broadened by Fermi motion), and $P_{\pi^0}$, P$_{\pi^{\pm}}$ are the 
four-momenta of the pions. Spectra for two typical energy regions are shown in Fig.~\ref{fig:mismas}.
These spectra were measured with the solid butanol target.
As discussed in Ref.~\cite{Dieterle_17}, the background contributions from the unpolarized heavier 
nuclei with large Fermi momenta smear out all structures. Therefore, only the data after subtraction
of the carbon background are shown. For the lower energy range, where the asymmetry is sufficiently large, the count rate difference for the two polarization states with parallel ($(\uparrow\uparrow)$) 
and antiparallel ($(\uparrow\downarrow)$) spin of photon beam and target are shown. For both spectra, 
only reactions off polarized deuterium nuclei contribute. The difference between the two types of spectra 
for the lower energy in the region of large missing mass is due to cancellation of background in 
the helicity difference spectra. The experimental data are compared to the line shapes of MC simulations 
for the $\gamma N\rightarrow N\pi^0\pi^{\pm}$ reaction and for background from triple-pion production, 
both for quasi-free nucleons bound in the deuteron (i.e. including Fermi motion constructed from the 
deuteron wave function \cite{Lacombe_81}). The agreement between the simulated line shapes and the 
experimental results is good. The indicated cuts were applied for further analysis. It should be noted that although the spectra shown in Fig.~\ref{fig:mismas} are  integrated over all angles and large bins of 
incident photon energies, the detailed analysis was done in finer energy bins and took into account the  dependence on cm angles of the pion pairs. The MC background structures are only roughly scaled to the background data, 
but show that their expected contribution to the signal region is small.

\subsection{Extraction of asymmetries and helicity-dependent cross sections} 

The asymmetry $E$ (or in notation for double pion production $P_z^{\odot}$ \cite{Roberts_05}) is 
defined by Eq.~\ref{eq:e}. It follows in principle directly from the count rates of the reaction for 
the two helicity states 
$(\uparrow\uparrow)$ ($N_{3/2}$) and $(\uparrow\downarrow)$ ($N_{1/2}$) by 
\begin{equation}
E=\frac{1}{P_{\odot}P_{T}}\cdot\frac{N_{1/2}-N_{3/2}}{(N_{1/2}-N_B)+(N_{3/2}-N_B)}~,
\label{eq:e_cr}
\end{equation}
where $P_{\odot}$ and $P_{T}$ are the beam and target polarization degrees and $N_B$ is the background
count rate from unpolarized nucleons bound in the heavy nuclei of the butanol molecules which cancels in
the numerator. Due to this background, count rate measurements with carbon and/or liquid deuteron targets 
are also needed. The asymmetry can then be constructed in two different ways. Equation~\ref{eq:e_cr} requires a measurement of the count rates $N_b$ with a carbon target. For this purpose, carbon foam
targets have been used, which had the same carbon surface density and the same geometry as the butanol 
target so that only the incident photon flux for both measurements had to be considered, all other 
systematical effects cancel or have no notable influence on final results. The other approach is to use in the denominator the results from a measurement 
with an unpolarized liquid deuterium target. In that case, fully normalized cross sections taking into 
account target density and detection efficiencies (can be slightly different for butanol and deuterium 
target because of different target geometry) have to be taken into account and the asymmetry can be 
expressed as the right hand side of Eq.~\ref{eq:e}. A comparison of the two methods was used 
to estimate systematic uncertainties due to the unpolarized background. The results are labeled 
`version 1' for normalization to deuterium data and `version 2' for subtraction of carbon background.   

Measurements with nucleons bound in deuterium are affected by the Fermi motion of the particles. 
This effect is mostly important when rapidly varying observables are smeared out in energy.
It appears when the tagged photon energy in the initial state is used to reconstruct 
the invariant mass $W$ of the final state nucleon - meson ensembles, because in this case the 
initial-state nucleon is approximated at rest. Alternatively, one can determine the final-state energy $W$
from the four vectors of the final state particle (defined as $W$ = $\sqrt{M^{2} + 2ME_{\gamma}}$, where $M$ is the target nucleon mass), which are however measured with less experimental resolution than the energy of the incident photon. This final state technique has been successfully applied in Refs.~\cite{Krusche_11,Dieterle_17,Witthauer_16,Witthauer_17,Kaeser_18,Dieterle_20,Werthmueller_14}, using the measured energies and angles of photons from $\pi^{0}$ or $\eta$ decays as well as the angle of the recoil nucleon. 

The difficulty in the present case occurs at higher energies because the kinetic energies of the charged pions are less well determined than those of photons 
due to punch-through of the particles, so that the resolution becomes poor at high incident photon energies. 

The most important systematic uncertainties for the asymmetry $E$ arise from the polarization degrees 
of the photon beam $P_{\odot}$ and the target $P_{T}$. As discussed for the analysis of other reactions 
from the same data set, they are estimated at $\Delta P_{\odot}=2.7$\% and $\Delta P_{T}=10$\%
\cite{Dieterle_17,Witthauer_17,Kaeser_18,Dieterle_20}. The relatively large uncertainty of the target
polarization was due to problems with the homogeneity of the magnetic field of the polarizing magnet, as
discussed in detail in \cite{Witthauer_17}.

Uncertainties from target densities, photon fluxes, and simulated detection efficiencies cancel in first 
order in the asymmetry $E$. Only the contributions from unpolarized background in the denominator can 
cause higher order effects, either because the carbon background has to be subtracted or the normalization 
is carried out with the liquid deuterium results, so that systematic differences between the measurements
(from photon flux, target density, or in case of the deuterium target non identical target geometry) can 
contribute. They are difficult to quantify and the best estimate comes from the comparison of the two
independent normalization methods. For both methods, only count rates normalized to fluxes and detection
efficiencies have been used, so that effects from different target geometries were removed at the 
level of precision of the MC simulations.      
   
For the helicity dependent absolute cross sections $\sigma_{1/2}$ and $\sigma_{3/2}$, the absolute uncertainties of fluxes, target densities, and detection efficiencies also contribute. Typical
uncertainties for target densities are $\approx$4\% and uncertainties of photon fluxes are $\approx$3\% 
range. In the present case, comparisons of the analysis of the results from the butanol and the liquid 
deuterium target help to estimate such effects. The uncertainty of the detection efficiency obtained from
the MC simulations is the most critical part. This is mainly due to the charged pions which could not 
be cleanly identified in the TAPS detector, so that at polar angels below $20^{\circ}$ charged pions were not accepted. This makes the MC simulations more prone to systematic effects from the
angular distribution of the charged pions. Such effects have been minimized by adapting the MC generator
to the measured data. In an iterative process, contributions from reaction chains such as $\gamma N\rightarrow \Delta^{\pm}\pi^0\rightarrow N\pi^{\pm}\pi^0$, 
$\gamma N\rightarrow \Delta^{0}\pi^{\pm}\rightarrow N\pi^{\pm}\pi^0$,
$\gamma N\rightarrow N\rho^{\pm}\rightarrow N\pi^{\pm}\pi^0$,
and phase space $\gamma N\rightarrow N\pi^{\pm}\pi^0$ (which subsumes many other contributions) have been fitted to the angular distributions of the pions and the invariant mass distributions of pion-pion and pion-nucleon pairs. For the absolute cross sections, the data from the liquid deuterium target were used.
    
\section{Results and Conclusions}
\label{sec:results}

\begin{figure}[thb]
\centerline{\resizebox{0.5\textwidth}{!}{
\includegraphics[width=0.45\textwidth]{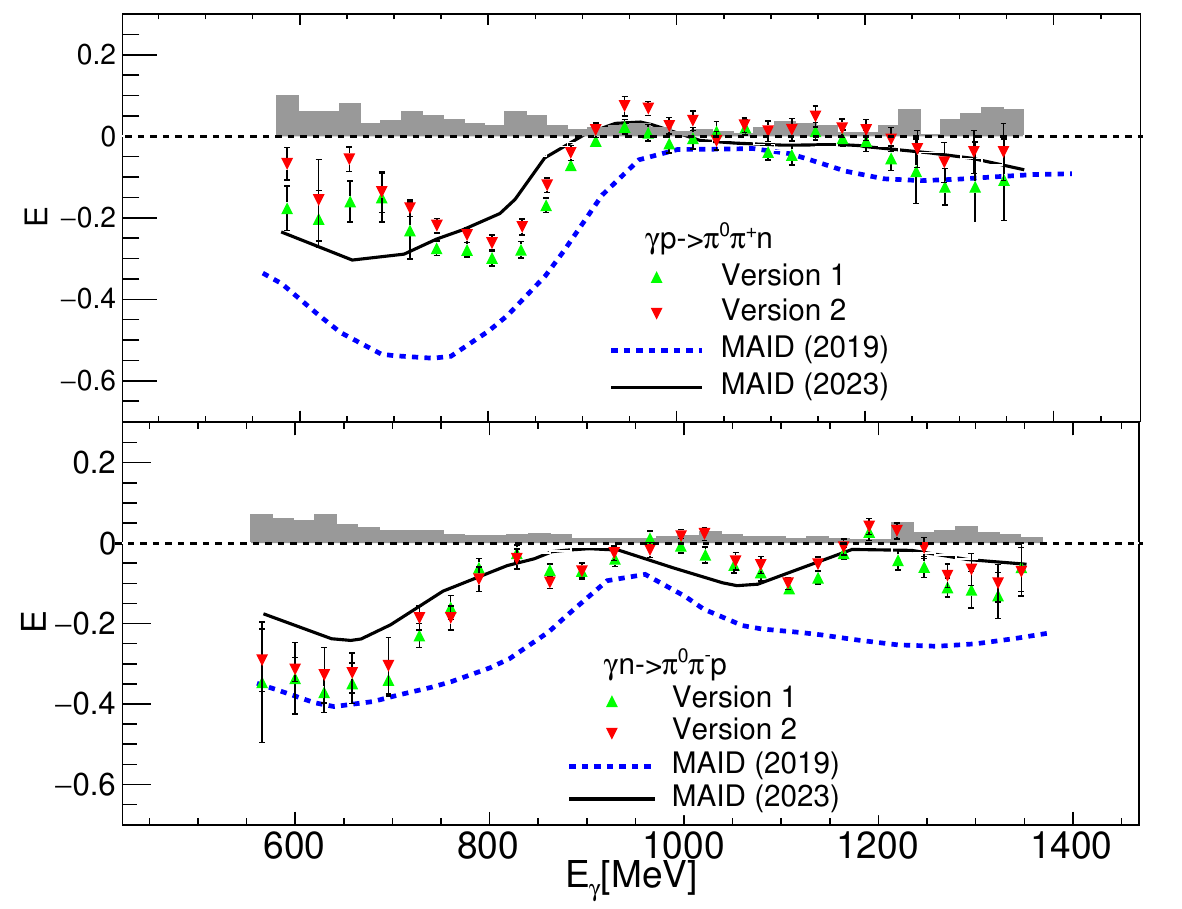}
}}
\caption{The asymmetry $E$ for the $\gamma p\rightarrow \pi^{0} \pi^{+}n$ (upper plot) and 
  $\gamma n \rightarrow \pi^{0} \pi^{-}p$ (lower plot) reactions as function of incident
  photon energy. 
  (Red) downward triangles: carbon-background subtraction, (green) upward triangles: normalization to 
  liquid deuterium data.
  Histograms: systematic uncertainty. (Blue) dotted and (black) solid lines are results from the MAID 
  model.
  The most recent version (MAID 2023) includes results from the present measurement in the data base.}
\label{fig:eobs}
\end{figure}

The results obtained for the total asymmetry $E$, integrated over all meson angles and invariant mass distributions, 
are shown in Fig.~\ref{fig:eobs} for both reaction channels, as function of the incident photon energy $E_{\gamma}$. 

\begin{figure*}[hbtp!]
  \centering 
 {\includegraphics[width=0.45\textwidth]{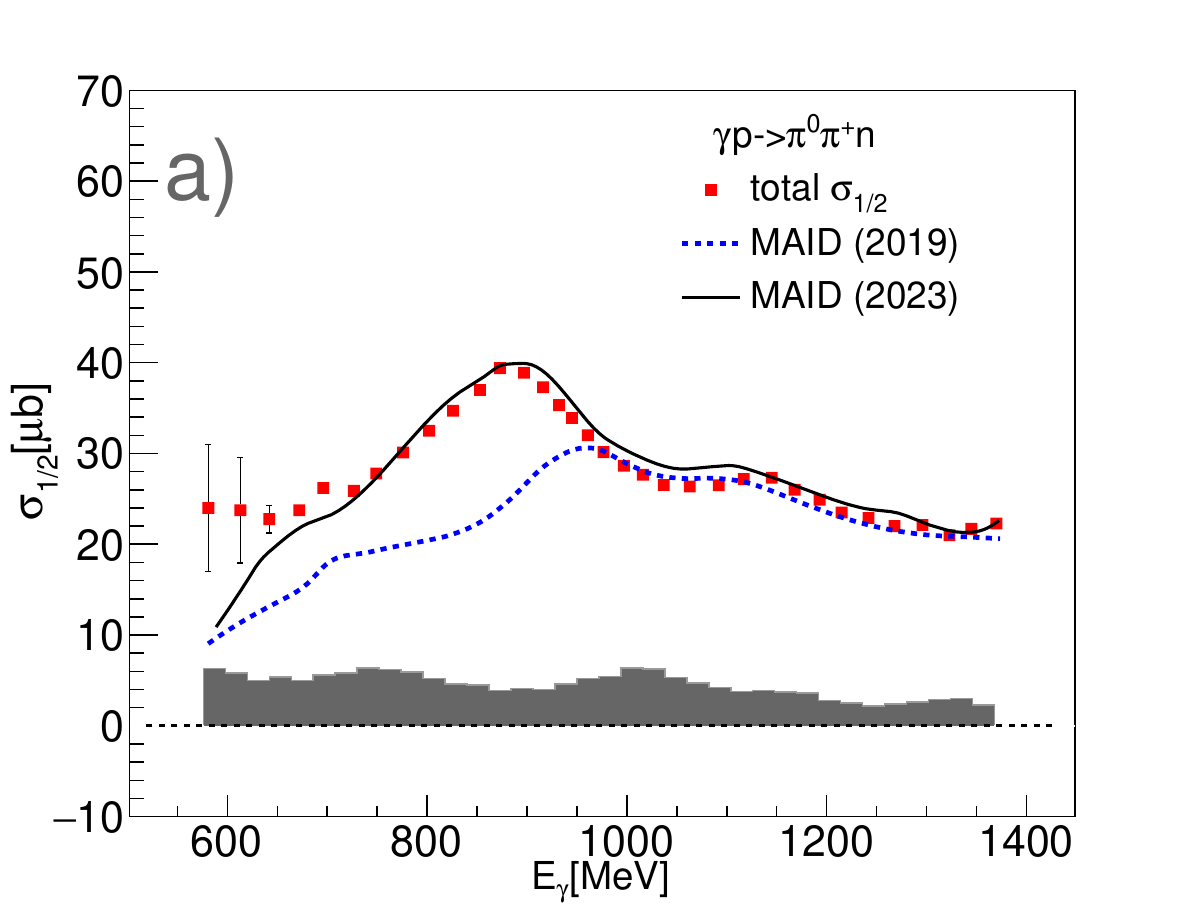}}
 {\includegraphics[width=0.45\textwidth]{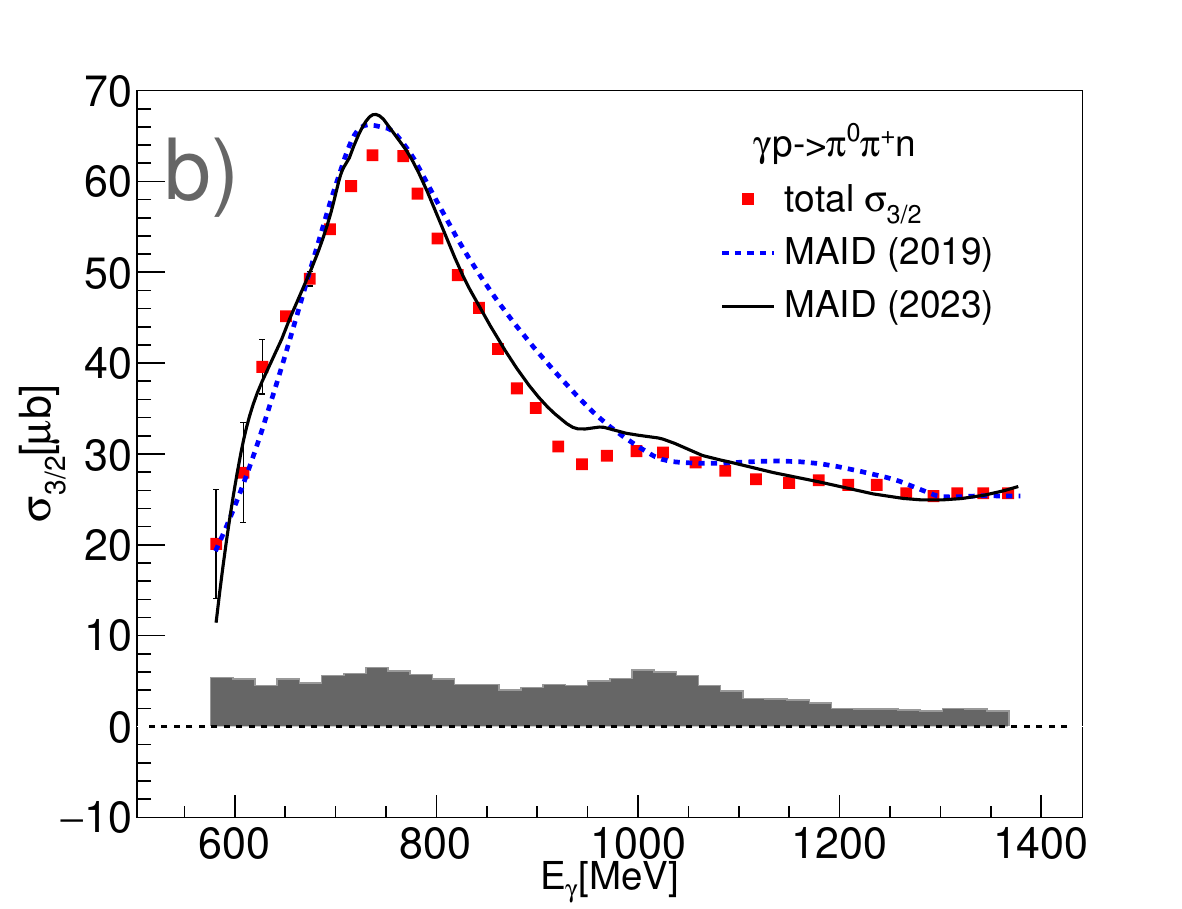}}\\
 {\includegraphics[width=0.45\textwidth]{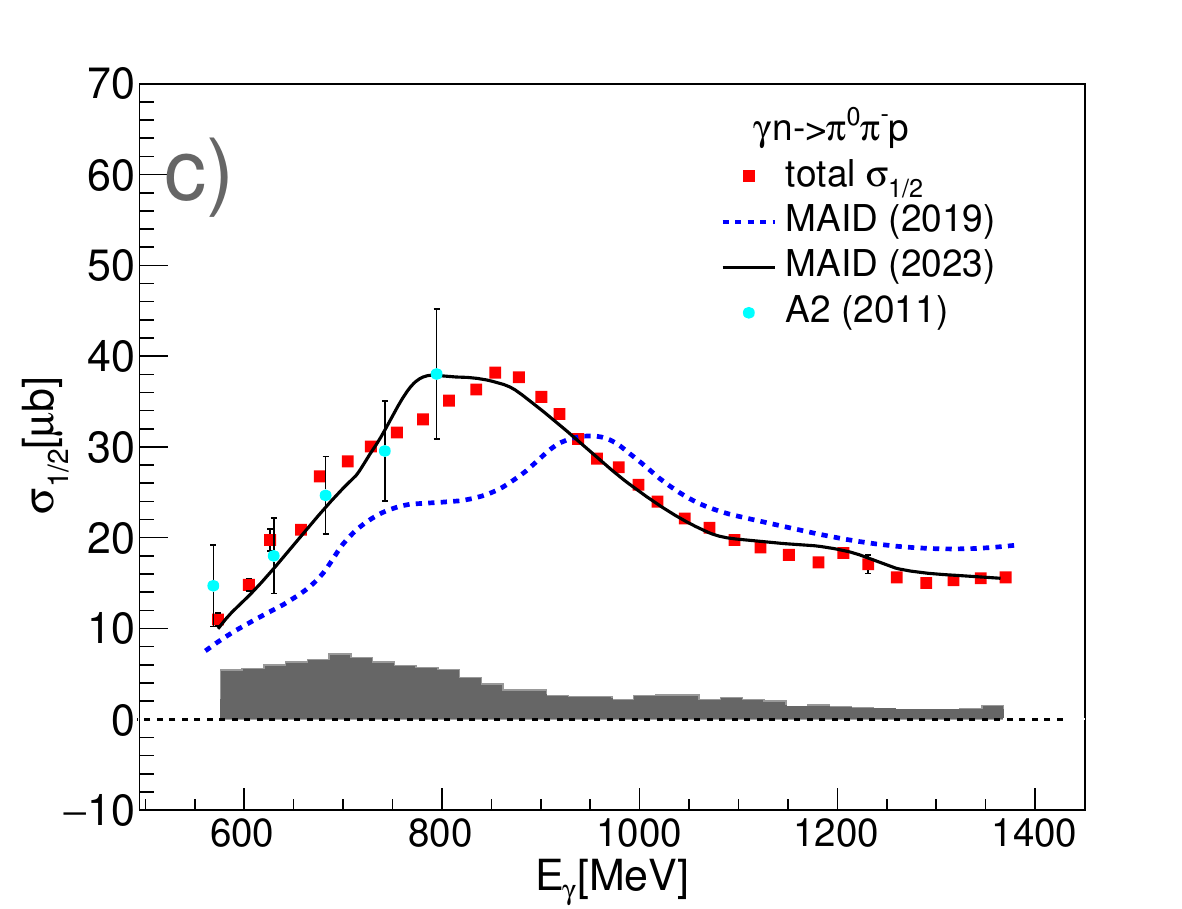}}
 {\includegraphics[width=0.45\textwidth]{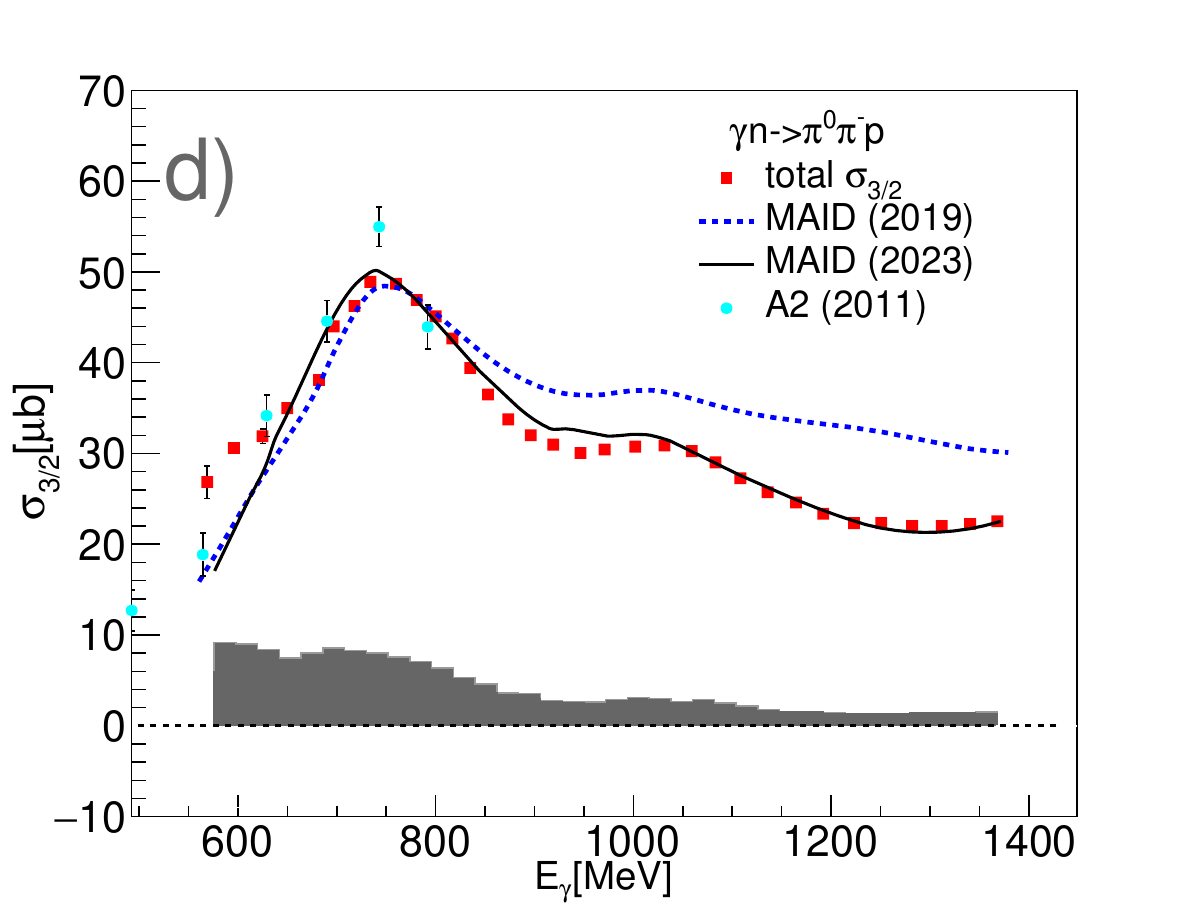}}
  \caption{Total cross sections for the two helicity states $\sigma_{1/2}$ (plots a,c) and $\sigma_{3/2}$ 
  (plots b,d) as function of incident photon energy $E_{\gamma}$ for the two reactions $\gamma p\rightarrow \pi^0\pi^+n$ (plots a,b) and $\gamma p\rightarrow \pi^0\pi^+n$ (plots c,d). 
  Red squares: our new data, cyan circles: existing GDH/A2 data~\cite{Ahrens_11}, dashed (blue) and solid (black) curves: results from the MAID model. 
  The gray bands indicate the systematic uncertainty. }
  \label{fig:pocs}
\end{figure*}

Figure~\ref{fig:eobs} summarizes the results from both analyses, where either in the denominator of Eq.~\ref{eq:e} the unpolarized background measured with
a carbon target is subtracted or the data are normalized to the unpolarized cross section measured with a 
liquid deuterium target. Both analyses agree very well, demonstrating that the elimination of the unpolarized
background is well under control. For the further analysis of the $\sigma_{1/2}$ and $\sigma_{3/2}$ 
components of the cross section the average of the two results for $E$ were used. The statistical
uncertainties of $E$ were linearly averaged because they are dominated by the fluctuations of the numerator
in Eq.~\ref{eq:e}, which is identical for both analyses. 

Figure~\ref{fig:pocs} shows the total cross sections for proton and neutron targets as function $E_{\gamma}$. They are compared to fits
with the MAID model. The new data, being in good agreement with the existing data from DAPHNE for the channel  $\gamma n\to p\pi^-\pi^0$~\cite{Ahrens_11}, provide a significant improvement in quality at energies up to $\approx$~800~MeV and extend the coverage towards higher energies.

The model used for fitting the data was obtained by refining the $2\pi-$MAID model of Ref.~\cite{Fix_05}. Its previous version, hereinafter referred to as MAID(2019), in which the present data were not included, was partially used in Ref.~\cite{Dieterle_20} to study partial wave content of the helicity components $\sigma_{1/2}$ and $\sigma_{3/2}$ of the $\gamma p\to p\pi^0\pi^0$ and $\gamma n\to n\pi^0\pi^0$ cross sections. In contrast to the original model of \cite{Fix_05}, where for the resonance parameters (masses, hadronic and radiative widths) the mean values from the Particle Data Group compilation were taken, in the present study these parameters were fitted to the data by varying them in a certain narrow range around their mean values given in Ref.~\cite{PDG_20}. Another difference from Ref.~\cite{Fix_05} is that the new $\gamma N\to \pi\pi N$ amplitude contains additional background terms in the $\pi\Delta$ channel in the partial waves with spin-parity $J^\pi=\frac32^\pm$ and isospin $I=\frac12$. The main intent of introducing these pure phenomenological terms was to eliminate the principal disagreement between the theory and the data  in the $\pi^0\pi^0$ channel in the energy region below the $N(1520)\frac32^-$ resonance. The major constraint of the new model is that the background amplitudes are smooth functions of energy compared to strong variation of the amplitudes in the resonance sector.

The FSI effects appearing in $\pi\pi$ production on a deuteron were calculated in Ref.~\cite{Fix_05}. They have been shown to be comparable or smaller than the systematic uncertainties of our new data on the cross sections for the reactions $\gamma p\to n \pi^0\pi^+$ and $\gamma n\to p \pi^0\pi^-$. The FSI effects in the $E$ asymmetry (defined as a ratio of the cross sections) are expected to be even smaller. In this regard, the model calculations presented here were performed on free nucleons.

\begin{figure*}
    \centering

{\includegraphics[width=0.45\textwidth]{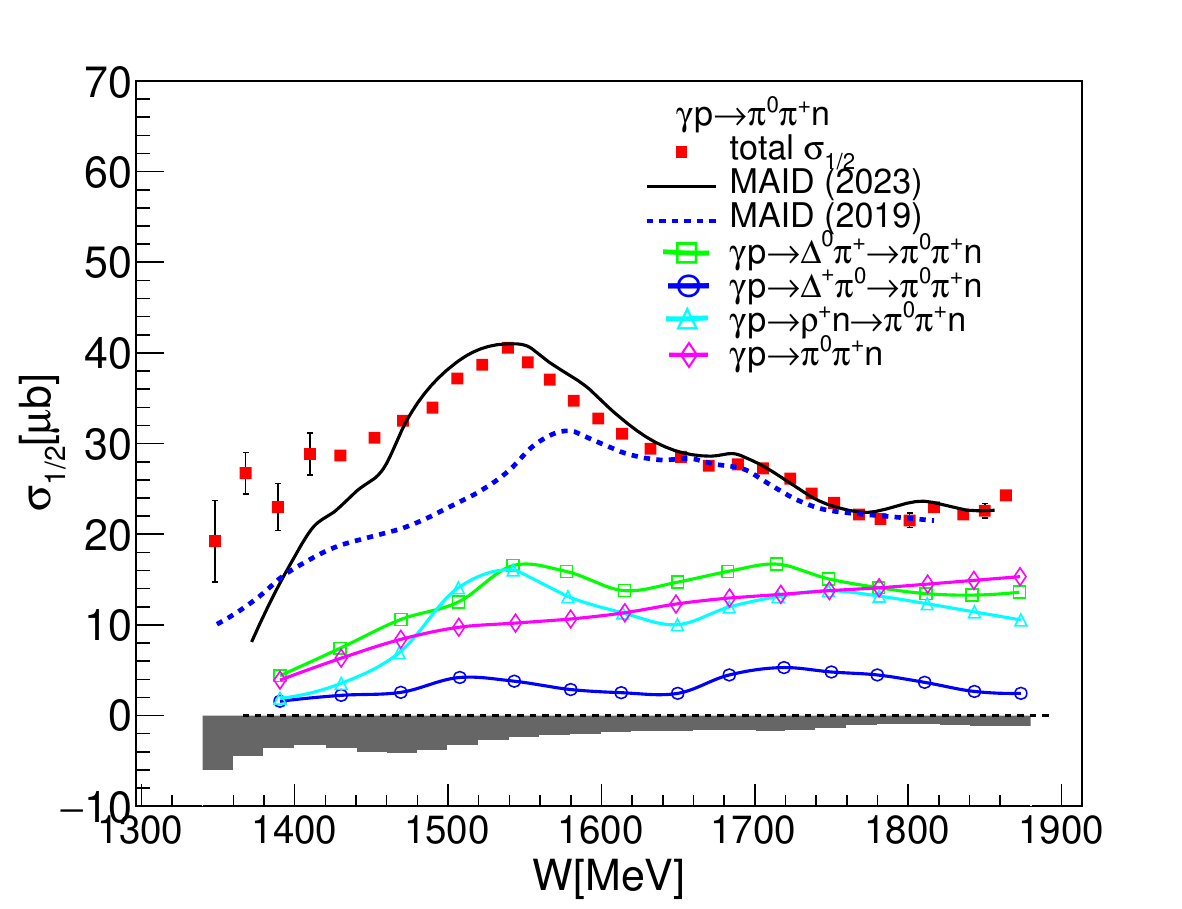}}
{\includegraphics[width=0.45\textwidth]{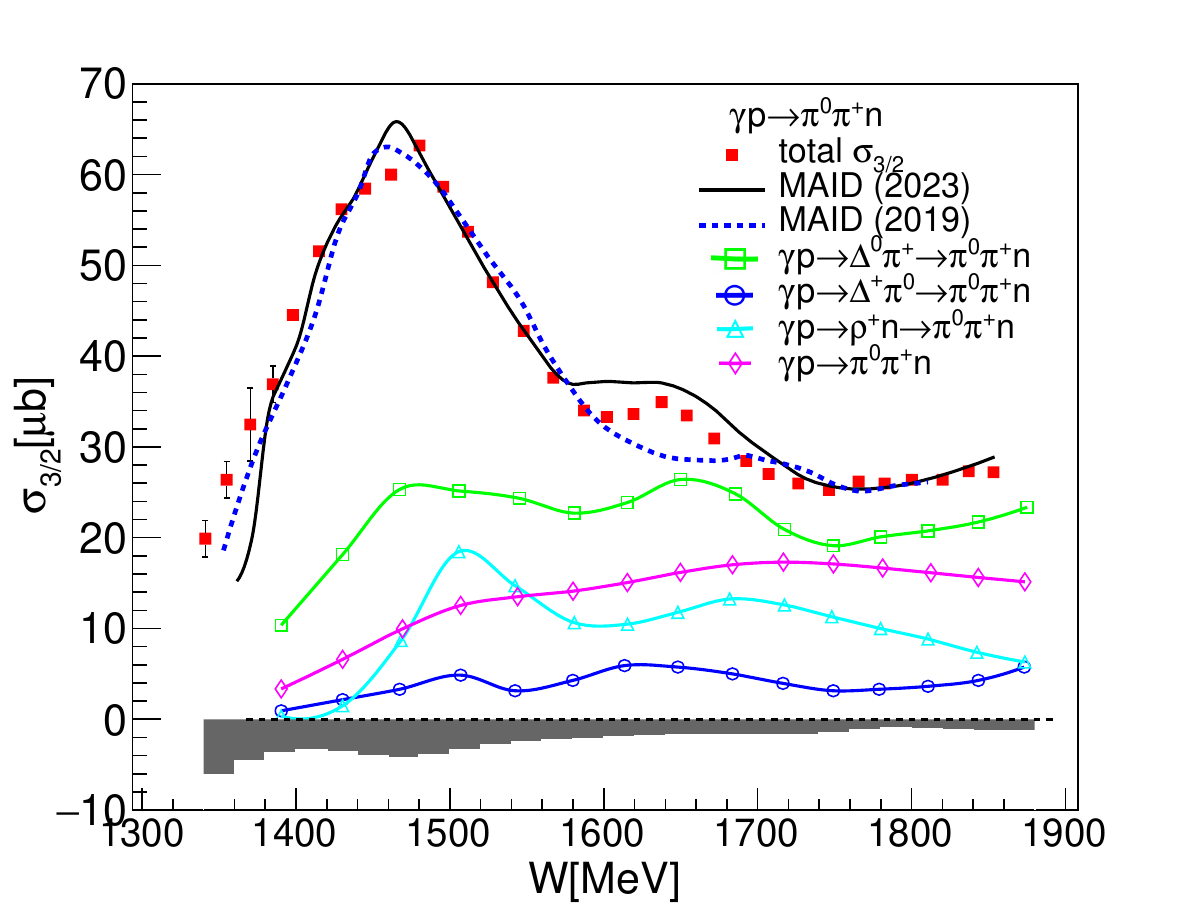}}\\
{\includegraphics[width=0.45\textwidth]{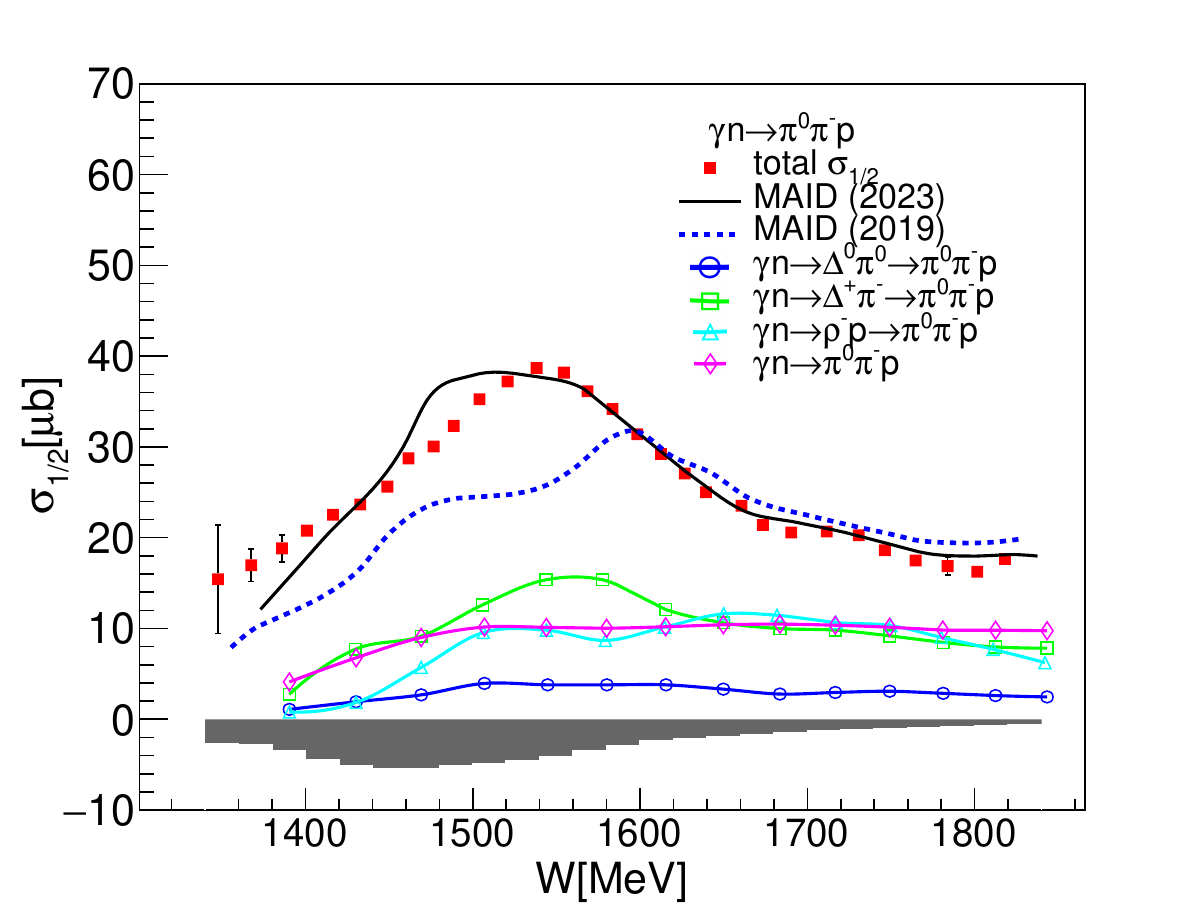}}
{\includegraphics[width=0.45\textwidth]{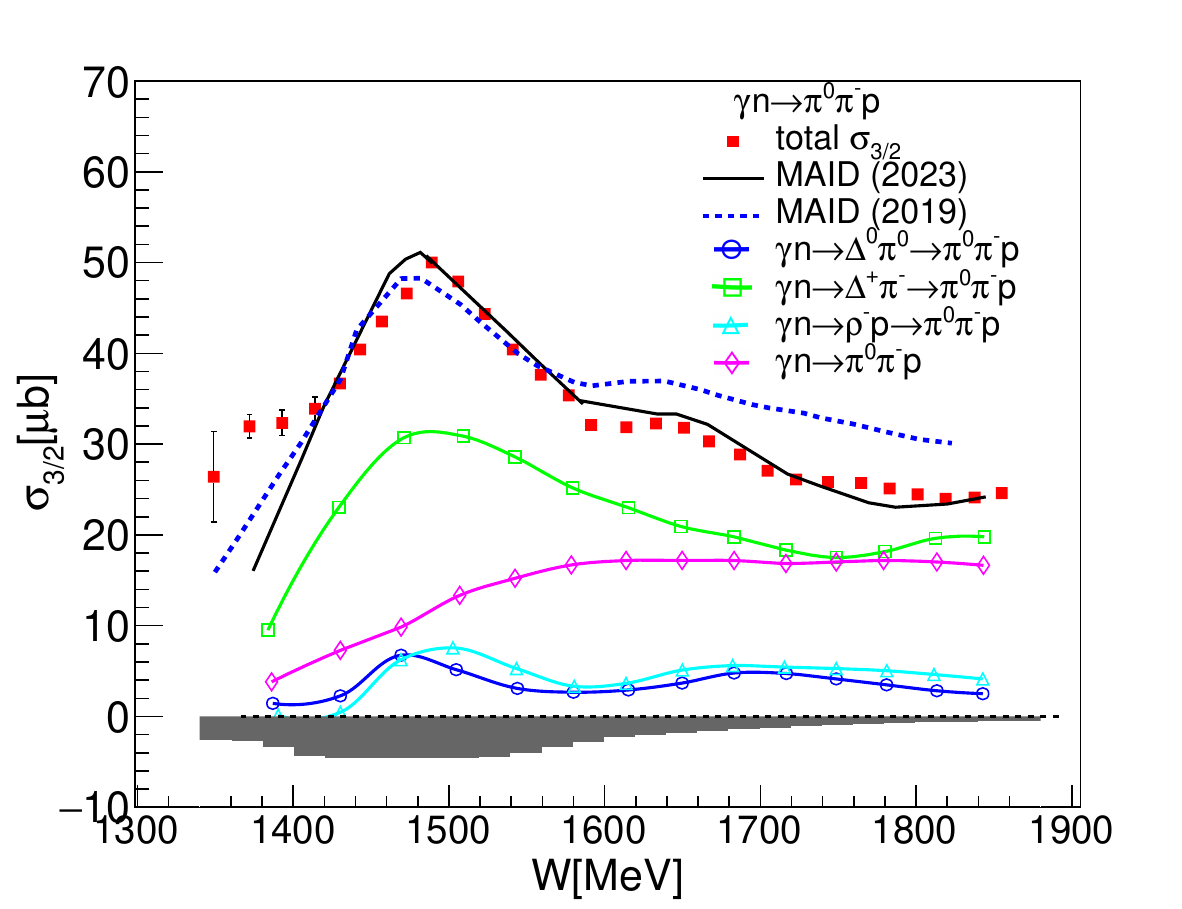}}
  
  \caption{Comparison of MAID model calculations (including individual reaction channels) with the experimentally measured total cross section.}
    \label{fig:foobar}
\end{figure*}

Figure~\ref{fig:foobar} summarizes the split of the total cross sections 
into different reaction channels for proton and neutron target and both helicity components. The study of individual mechanisms in the $\pi^\pm\pi^0$ production~\cite{Gomez_96,Ochi_97,Fix_05} shows that in the second resonance region the main contribution to these channels comes from the $\Delta$ Kroll-Ruderman term, as well as from the excitation of the $N(1520)3/2^-$ resonance. The Kroll-Ruderman mechanism dominates in the region of photon energies $E_\gamma < 0.6$ GeV and leads to the production of a $\pi^\pm\Delta^0$ configuration with the orbital momentum $l=0$. Thus, the relative contribution of this term to $\sigma_{1/2}$ and $\sigma_{3/2}$ is basically determined by the ratio of the squared Clebsch-Gordan coefficients $(\frac{1}{2}\,\frac{1}{2}\,11|\frac{3}{2}\,\frac{3}{2})$ and $(\frac{1}{2}\,-\frac{1}{2}\,11|\frac{3}{2}\,\frac{1}{2})$, which gives $\sigma_{3/2}=3\sigma_{1/2}$. As one can see from Fig.~\ref{fig:pocs}, this simple relation is somewhat violated due to admixture of other terms.

The rapid change of $\sigma _{3/2}$ in the $\pi\Delta$ and $\rho N$ channels at the energies around $E_\gamma=770$ MeV can naturally be attributed to the excitation of $N(1520)3/2^-$. The photon decay amplitude $A_{\lambda}$ of this resonance is significantly larger for the helicity state $\lambda=3/2$ than for $\lambda=1/2$ for both the proton and neutron target \cite{PDG_20}, so that this state is primarily seen in the $\sigma_{3/2}$ part. This applies to a large extent to the $\rho N$ channel. According to various model calculations \cite{Gomez_96,Ochi_97,Fix_05}, the background mechanisms of the $\rho N$ production in the second resonance region are rather insignificant, so that this channel should mostly be saturated by the $\gamma N\to N(1520)\to \rho N$ transition. This assumption gives for the ratio of the corresponding partial cross sections $R=\sigma^{(\rho N)}_{3/2}(\pi^+\pi^0)/\sigma^{(\rho N)}_{3/2}(\pi^-\pi^0)\approx[A^{(p)}_{3/2}/A^{(n)}_{3/2}]^2$. Taking $A^{(p)}_{3/2}=0.140$\,GeV$^{-1/2}$ and $A^{(n)}_{3/2}=-0.115$\,GeV$^{-1/2}$ from \cite{PDG_20} we obtain $R=1.48$ in reasonable agreement with our data. The fit gives somewhat larger value, $R\approx 2.25$. This difference manifests itself in the underestimation of the cross section $\sigma_{3/2}^{(\rho N)}$ in the $\pi^-\pi^0 p$ channel.

As for contribution of other resonances, from simple considerations based on the isospin algebra, one concludes 
that if the two pions are produced by excitation of an isospin $I=\frac12$ resonance, $\gamma N\to N^*\to\pi\Delta\to \pi\pi N$, the formation of intermediate states with charged and neutral pions, $\pi^\pm\Delta$ and $\pi^0\Delta$, has the same probability. That is, $\sigma^{(\pi^\pm\Delta)} = \sigma^{(\pi^0\Delta)}$ for the $N^*$-type resonances. For the $\Delta^{\star}$ states, using the same simple considerations, one obtains $\sigma^{(\pi^\pm\Delta)}= 4 \sigma^{(\pi^0\Delta)}$. For this reason, in the second resonance region, where the $I=1/2$ resonances are mainly excited, the visible excess of the $\pi\Delta$ contribution with the charged pions ($\pi^ +\Delta^0$ and $\pi^-\Delta^+$) over the corresponding neutral pion channels ($\pi^0\Delta^+$ and $\pi^0\Delta^0$) should primarily be related to the Born mechanisms.

Finally, we note the difference between MAID(2019) and MAID(2023) versions of the $2\pi$-MAID model. As is noted above, in MAID(2019) the present results were not included, and the major part of the fitted database was provided by numerous experimental results for the $\pi^0\pi^0 p$ and $\pi^0\pi^0 n$ channels. Inclusion of the present data into the fitting procedure in MAID(2023) led primarily to a change of hadronic partial widths of some resonances. In particular, to reproduce $\sigma_{1/2}$ at energies below $E_\gamma = 800$ MeV, the $\pi\Delta$ and $\rho N$ branching ratios of $N(1535)1/2^-$ were increased from $1\,\%$ and $5\,\%$ to $7\,\%$ and $16\,\%$, respectively. At higher energies, $E_\gamma > 1$ GeV, we had to slightly reduce the contributions of the resonances $\Delta(1620)1/2^-$ and $\Delta(1700)3/2^-$. Here also the diffractive $\rho$ production via meson exchange in the $t$-channel was additionally suppressed by artificial amplifying the absorptive corrections.

In summary, precise data have been obtained for the helicity dependence of photoproduction of mixed-charge pion pairs off nucleons. This letter also summarizes the total cross sections. Angular distributions and invariant-mass distributions of the pion-pion and pion-nucleon pairs will be discussed in a more detailed
comparison of the experimental data to the reaction model results in an upcoming paper.

\vspace*{0.5cm}
{\bf Acknowledgments}\\

This paper is dedicated to the memory of B. Krusche.\\

We wish to acknowledge the outstanding support of the accelerator group and operators of MAMI.
This work was supported by Schweizerischer Nationalfonds (200020-156983, 132799, 121781, 117601),
Deutsche For\-schungs\-ge\-mein\-schaft (SFB 443, SFB 1044, SFB/TR16), the INFN-Italy,
the European Community-Research Infrastructure Activity under FP7 programme (Hadron Physics,
grant agreement No. 227431), the UK Science and Technology Facilities Council 
(ST/J000175/1, ST/G008604/1, ST/G008582/1,ST/J00006X/1, and ST/L00478X/1),
the Natural Sciences and Engineering Research Council (NSERC, FRN: SAPPJ-2015-00023), Canada. 
The model calculation was supported by 
the Russian Science Foundation, Grant No. 22-42-04401.
This material is based upon work also supported by the U.S. Department of Energy, Office of Science, Office of Nuclear Physics Research Division, under Award Numbers DE-FG02-99-ER41110, DE-FG02-88ER40415, DE-FG02-01-ER41194, and DE-SC0014323 and by the National Science Foundation, under Grant Nos. PHY-1039130 and IIA-1358175.

\end{document}